\documentclass[fleqn,usenatbib]{mnras}
\usepackage{graphicx}\usepackage{epsfig}\usepackage{epsf}
\usepackage{amsmath}\usepackage{amssymb}\usepackage{stfloats}
\usepackage{soul}
\usepackage[usenames]{color}
\usepackage{tree-dvips}

\definecolor{green}{rgb}{0.3,0.7,0.}
\newcommand{\ra}{\text{RA}}
\newcommand{\dec}{\text{Dec}}
\title[Gaia Wd1]{Inferring the parallax of Westerlund 1 from {\it\textbf{Gaia}} DR2}

\author[M. Aghakhanloo et al.]{Mojgan Aghakhanloo$^{1}$\thanks{E-mail:
ma14g@my.fsu.edu}, Jeremiah W. Murphy$^1$, Nathan Smith$^2$, John Parejko$^3$,\newauthor Mariangelly D\'iaz-Rodr\'iguez$^1$, Maria R. Drout$^{4}$\thanks{Hubble and Carnegie-Dunlap Fellow}, Jose H. Groh$^6$, Joseph Guzman $^1$ \newauthor and  Keivan G. Stassun$^{7,8}$\\$^1$Department of Physics, Florida State University, 77 Chieftan
Way, Tallahassee, FL 32306, USA\\$^2$Steward Observatory, University of Arizona, 933 N. Cherry Ave., Tucson, AZ 85721, USA\\$^3$Department of Astronomy, University of Washington, Box 351580, Seattle, WA 98195, USA\\$^4$The 
Observatories of the Carnegie Institution for Science, 813 Santa Barbara St, Pasadena, 
CA 91101, USA\\$^5$School of Physics, Trinity College Dublin, The University of Dublin, Dublin, Ireland\\$^6$Department of Physics \& Astronomy, Vanderbilt University, 6301 Stevenson Center Lane,
Nashville, TN 37235, USA \\$^7$Department of
Physics, Fisk University, 1000 17th Avenue N., Nashville, TN 37208, USA}

\begin{document}
\pagerange{\pageref{firstpage}--\pageref{lastpage}} \pubyear{2019}
\maketitle
\label{firstpage}

\begin{abstract}
Westerlund 1 (Wd1) is potentially the largest star cluster in the Galaxy. That designation critically depends upon the distance to the cluster, yet the cluster is highly obscured, making luminosity-based distance estimates difficult.  Using {\it Gaia} Data Release 2 (DR2) parallaxes and Bayesian inference, we infer a parallax of $0.35^{+0.07}_{-0.06}$ mas corresponding to a distance of $2.6^{+0.6}_{-0.4}$ kpc. To leverage the combined statistics of all stars in the direction of Wd1, we derive the Bayesian model for a cluster of stars hidden among Galactic field stars; this model includes the  parallax zero-point. Previous estimates for the distance to Wd1 ranged from 1.0 to 5.5 kpc, although values around 5 kpc have usually been adopted.  The {\it Gaia} DR2 parallaxes reduce the uncertainty from a factor of 3 to 18\% and rules out the most often quoted value of 5 kpc with 99\% confidence.  This new distance allows for more accurate mass and age determinations for the stars in Wd1. For example, the previously inferred initial mass at the main-sequence turn-off was around 40 M$_{\odot}$; the new {\it Gaia} DR2 distance shifts this down to about 22 M$_{\odot}$.  This has important implications for our understanding of the late stages of stellar evolution, including the initial mass of the magnetar and the LBV in Wd1. Similarly, the new distance suggests that the total cluster mass is about four times lower than previously calculated. 
\end{abstract}


\section{INTRODUCTION}\label{sec:intro}
Massive stars are a central focus of ongoing work in stellar evolution theory. Many gaps exist in our understanding of massive stars due to their rarity, short lifetimes, high fraction of interacting binaries, and imprecise Galactic distances. To better understand the evolutionary path of massive stars, it is helpful to explore associated clusters or OB associations. If one can show that the star is a member of a cluster or association, studying the cluster provides a unique insight into intrinsic properties of cluster members. 
For example, Westerlund 1 has a Luminous Blue Variable \citep[LBV;][]{CN04}, at least 24 Wolf\_Rayet stars \citep[WR;][]{C05,C06,groh06,F18}, 6 yellow hypergiants \citep[YHG;][]{C05}, and a magnetar \citep{M06}; knowing the distance to this one cluster will help to constrain the luminosity, mass, and evolution of all of these late phases of stellar evolution. However, our sightline to the cluster suffers from substantial extinction and reddening, which has made its distance difficult to estimate using luminosity indicators \citep{P98,C05,D16}. {\it Gaia} parallaxes provide an independent distance indicator.

The massive young star cluster, Westerlund 1 (Wd1), was detected by \citet{W61} during a survey of the Milky Way. Wd1 is located at $\rmn{RA}(2000)=16^{\rmn{h}} 47^{\rmn{m}} 04\fs0$, $\rmn{Dec.}(2000)=-45\degr 51\arcmin 04\farcs 9$, which corresponds to Galactic coordinates of $\ell=339.55\degr$ and 
$b=-00.40\degr$.
 
The first distance estimates mostly relied on reddening-distance relationships and ranged from 1.0 to 5 kpc \citep{W61,W68,P98,C05,C06}. \citet{W61} first suggested an extinction of $A_{V} = 12.0$ mag, and reported a distance of 1.4 kpc. Later, \citet{W68} derived a significantly larger distance of 5 kpc by using VRI photographic photometry with near-infrared photometry of the brightest stars. In contrast, \citet{P98} presented CCD imaging in the V and I bands, and using isochrone fitting, estimated a distance of $1.0\pm{0.4}$ kpc.

More recently, \citet{C05} obtained spectra for the brightest members of Wd1 and improved our knowledge of Wd1.  With more detailed spectra, many of the brightest members were identified as post-main-sequence stars. Since the isochrone fitting of \citet{P98} assumed that many of the stars are on the MS, the \citet{P98} distance estimate was incorrect. Six of the stars in \citet{C05} are YHGs, and the most luminous YHGs are presumed to have relatively standard luminosity of around  log$(L/L_{M_\odot}) \sim 5.7$ \citep{S04}. Assuming that the YHGs in Wd1 were at the observed upper luminosity limit for cool hypergiants, and adopting an extinction of $A_V=11.0$, \citet{C05} inferred a distance of $\la$ 5.5 kpc.  However, they noted that their reddening law is not entirely consistent with Wd1 data.  To place a lower limit on the distance, they noted a lack of radio emission from the WR winds; this suggests a minimum distance of $\sim$2 kpc. Hence, \citet{C05} reported a distance of $2< R <5.5$ kpc.
However, these constraints on the distance and reddening depend sensitively on assumptions of wind physics for evolved stars, which is still uncertain \citep{smith14}.

A variety of subsequent investigations produce similar results and accuracy.  \citet{C06} inferred a similar distance using near-IR classification of WN and WC stars. More recently, \citet{K07} derived a distance of 3.9 $\pm$ 0.7 kpc based on the radial velocity of H\textsc{i} features in the direction of Wd1. \citet{B08} studied the population of stars below $\sim$30 M$_{\odot}$ and derived a distance of 3.55 $\pm$ 0.17 kpc based on the apparent brightness and width of the CMD using Pre-MS isochrones. Later, \citet{G11} performed a similar analysis but also modelled the completeness, field star subtraction, and error propagation, and inferred a distance value of 4.0$\pm$0.2 kpc.  Many of the previous techniques require some assumption about the luminosity of spectral classes.  Recently, \citet{K12} used the dynamics and the geometry of an eclipsing binary (W13) to infer a distance of 3.7$\pm$0.6 kpc. Given all of difficulties explained above in estimating the distance, it is necessary to use {\it Gaia} Data Release 2 (DR2) to find an independent distance estimate to Wd1.

These distances have implications for the mass, age, and cluster members.  For example, assuming a distance of 5 kpc, \citet{CN04} inferred a total cluster mass of $\sim 10^5$ M$_{\odot}$ and an age of 3.5-5 Myr.  With this age, the magnetar's progenitor would have an initial mass of $>$40 M$_{\odot}$.  The distance is critical, but estimates of the distance suffer from large uncertainties and systematics. It is therefore valuable to infer the distance using a more accurate and geometric technique.

The main goal of this paper is to infer an independent and geometric distance to Wd1 using {\it Gaia} DR2. The {\it Gaia} DR2 parallax precision for individual stars in Wd1 ranges from 0.04 to 1.0 mas.  Since Wd1 is probably of order a few kpc in distance, the larger uncertainties prohibit a precise distance estimate for an individual star.  However, the combined statistics of all cluster members should easily produce a more precise distance. Section~\ref{sec:method} describes the data and method to infer the distance. First, we describe the {\it Gaia} DR2 data and possible systematic uncertainties. Then we infer the approximate parallax of the cluster by estimating the true mean parallax for several annuli from the cluster centre (Section~\ref{sec:analyticmodel}). In Section~\ref{sec:Bayes}, we use a Bayesian inference technique to infer the cluster parallax, cluster density, field-star density, the parallax zero-point of the cluster, field-star length scale, and the field-star offset. In Section~\ref{sec:results}, we present the inferred distance to Wd1 and the length scale for the field star distribution. Then, we compare this distance and field-star distribution with previous works (Section~\ref{sec:discussion}). We also discuss how the revised distance affects the properties of cluster members. Section~\ref{sec:consclusion} presents a summary and direction for further investigation. 
\section{Method}\label{sec:method}
In this section, we describe the method to infer the distance to Wd1. The stars in the direction of Wd1
comprise cluster stars as well as Galactic field stars.  Therefore, to infer the distance to Wd1, our likelihood model must account for both cluster and field stars.  The following sections describe the data and methods required to model both components and infer the distance to Wd1.

\subsection{{\it {\textbf{Gaia}}} Data Release 2 data}
\label{sec:data}
The source of the data is the {\it Gaia} DR2 \citep{G16,Ga18}. We collect all {\it Gaia} DR2 sources within 10 arcmin
of the position of Wd1; $\rmn{RA}(2000)=16^{\rmn{h}} 47^{\rmn{m}} 04\fs0$,
$\rmn{Dec.}(2000)=-45\degr 51\arcmin 04\farcs 9$. 

\begin{figure*}
\includegraphics[width=6.5in]{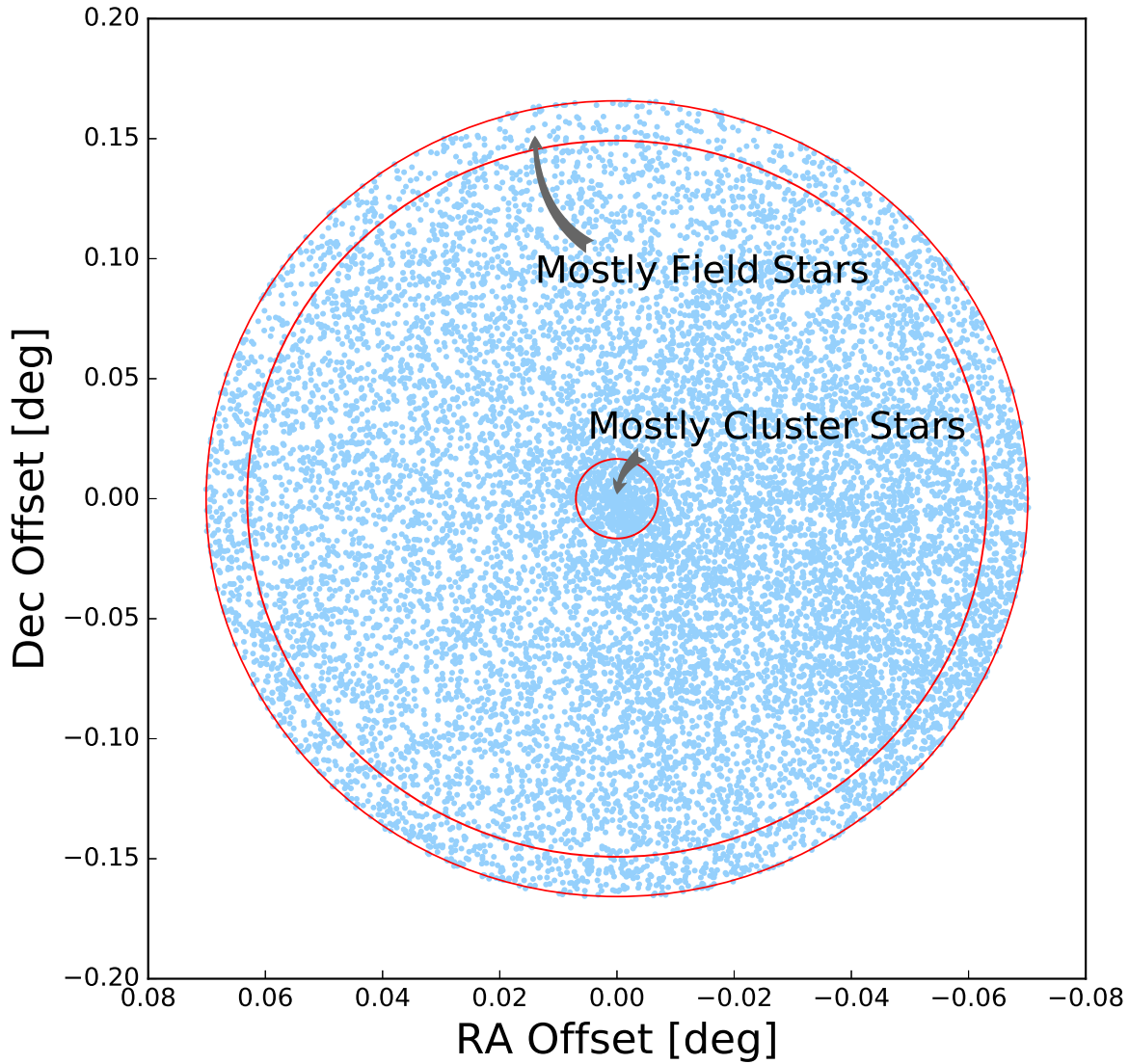}
\caption{Position of all {\it Gaia} stars within 10 arcmin of Westerlund 1. Axes are offsets from a given position in degrees on the sky; RA Offset=$(\ra-\ra_{\text{Wd1}})\cos(\dec._{\text{Wd1}})$ and Dec. Offset=$\dec.-\dec._{\text{Wd1}}$, where $\ra_{\text{Wd1}}=251.77\degr$ and $\dec._{\text{Wd1}}=-45.85\degr$. Stars in the inner circle, with 1 arcmin radius, are mostly associated with the cluster, while the stars in the outer ring are mostly field stars.}
\label{fig:Ring}
\end{figure*}

Fig.~\ref{fig:Ring} presents the positions of all objects within 10 arcmin of
Wd1. The inner circle marks a region that is 1 arcmin from the centre of the cluster. The outer annulus extends from 9 to 10 arcmin. The density of stars is mostly uniform throughout the field of
view, it is slightly over dense towards the right; however, the density does noticeably increase towards the
centre. This spatial separation between cluster and field
stars suggest a strategy for constraining the parameters for each
population. The inner circle contains both field and cluster stars,
but is dominated by cluster stars.  Therefore, the inner circle
provides a good constraint on the cluster population.  The outer annulus
are mostly field stars.  Therefore, the outer region will
constrain the field star distribution.

\begin{figure*}
\includegraphics[width=\textwidth]{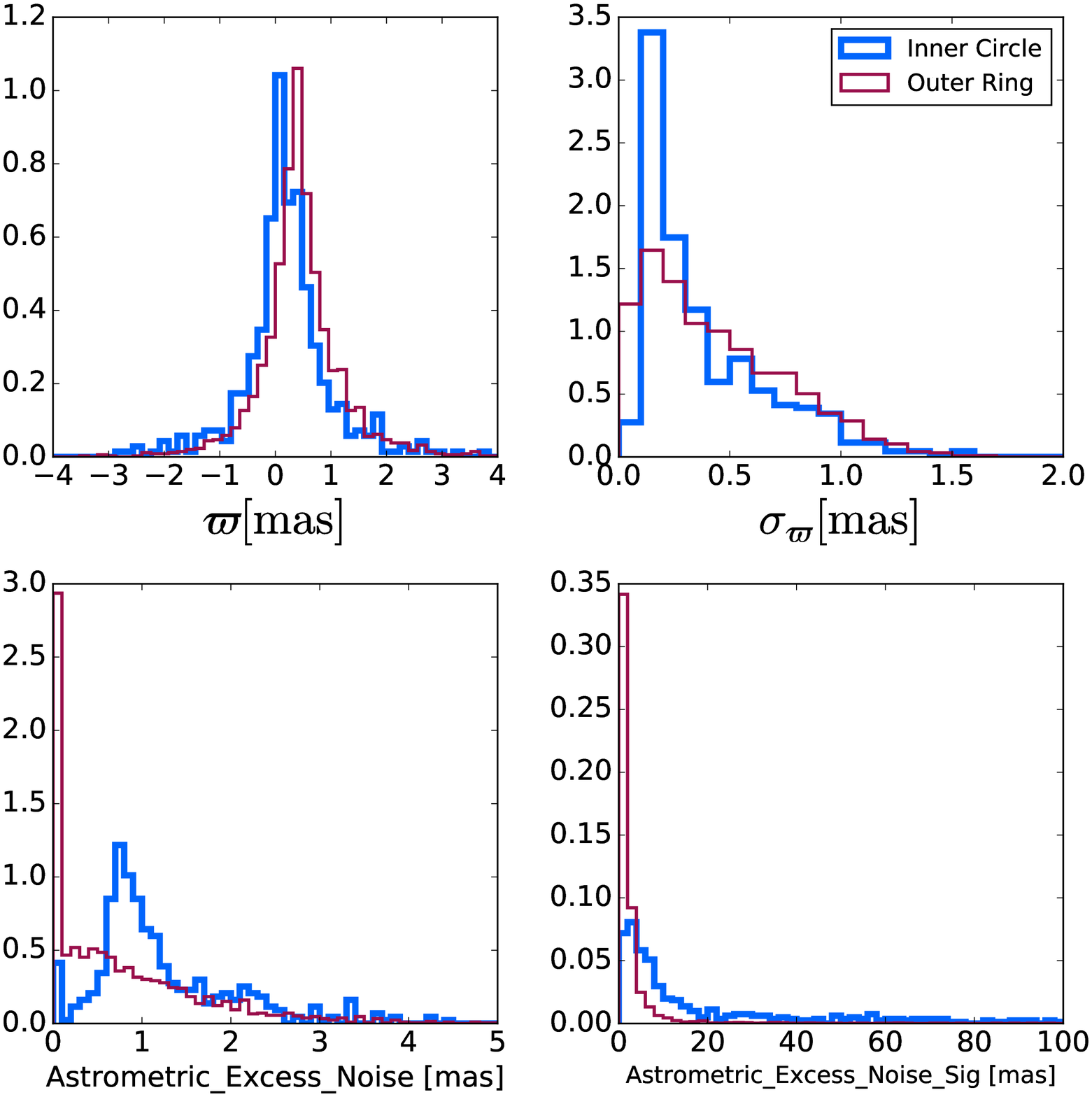}
\caption{Histograms of parallaxes (top left panel), parallax uncertainties (top right panel), astrometric excess noise ($\epsilon$, bottom left panel), and astrometric excess noise significance ($D$, bottom right panel) of all stars in the inner circle and the outer annulus. The average parallaxes for the inner circle and the outer ring are clearly distinct; the average parallax $\left <\varpi \right >=0.22\pm{0.04}$ mas for the inner circle, and $\left < \varpi \right > = 0.45\pm{0.02}$ mas for the outer ring.  A small fraction of the inner circle stars have small excess noise ($D < 2$), while the outer ring has a much larger fraction that satisfy this quality metric.}
\label{fig:Histalldata}
\end{figure*}

Fig.~\ref{fig:Histalldata} shows
  histograms of parallax, parallax uncertainty, astrometric excess
  noise, and astrometric excess noise significance of all stars in the inner
  circle and the outer annulus.  The average parallax for the inner
  circle is $0.22\pm{0.04}$ mas, and the average parallax for the
  outer annulus is $0.45\pm{0.02}$ mas. These averages include all stars, no filtering, and does not include the parallax zero-point. Never the less, the difference in average
  parallax already indicates that the cluster is farther than the
  average field star.  The distribution of expected parallax
  uncertainty (top right panel) is similar between the two regions. The minimum parallax uncertainty in the direction of Wd1 is 0.04 mas, but the vast majority of uncertainties are even larger (up to a few mas), including systematics such as parallax  zero-point \citep{L18}. The distribution for astrometric excess noise (bottom left panel)
  shows a considerable difference between the two regions. The astrometric excess noise ($\epsilon$) represents extra variation in the data that is not included in the five-parameter astrometric model; the astrometric noise significance is $D$ (bottom right panel). A higher
 fraction of objects in the inner circle have large excess noise. Therefore, we keep the sources with $D\leq2$ to avoid bad astrometric solutions. 

Although the astrometric excess noise and significance are useful for assessing the quality of the astrometric solutions, the renormalized unit error (RUWE) and visibility periods used are more useful in very crowded regions like Wd1.
The RUWE is the re-normalized goodness of fit ($\sqrt{\chi^2/(N-5)}$). $N$ is the total number of along scan observations used in the astrometric solution of the source. The high crowding is possibly responsible for the lower number of observations per star. A visibility period indicates the number of groups of observations separated from other groups by at least 4 d. Therefore, a higher number of visibility periods indicates that the solution is less vulnerable to errors. As recommended by \citet{L18}, we use astrometric solutions with at least eight visibility periods and RUWE $<1.40$. The astrometric excess noise, visibility periods, and RUWE cuts reduce the number of sources in the inner circle from 435 to 42 and reduce the number of sources in the outer ring from 2127 to 1344.

\begin{figure}
\includegraphics[width=0.5\textwidth]{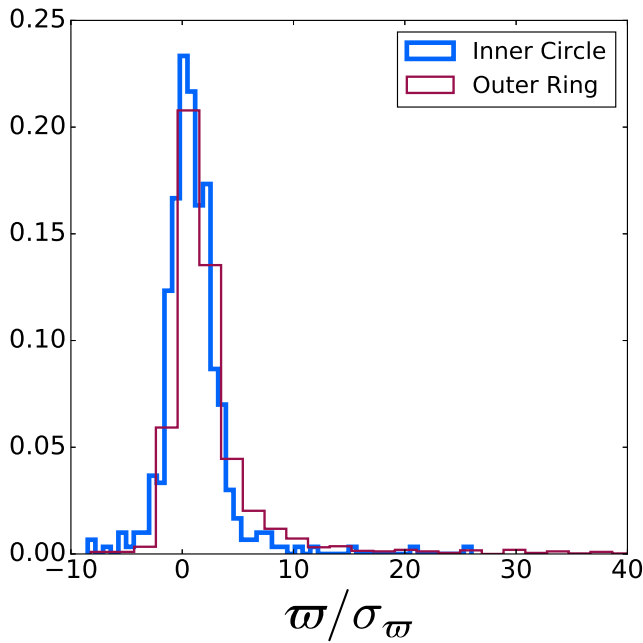}
\caption{Histogram of parallax over the uncertainty of all stars in the inner circle and the outer annulus. Many of the stars in the inner circle and the outer annulus have large uncertainties (ratios of order 1 or less). The large uncertainties and negative parallaxes indicate that inferring the distance to an individual star using parallax inversion $d=1/\varpi$ is either inaccurate or impossible.}
\label{fig:noise}
\end{figure}

If the uncertainties are relatively small, then direct inversion of the parallax is a reasonable inference of the distance, $d=1/\varpi$. However, Fig.~\ref{fig:noise} shows that most of the stars in the inner and the outer region have large uncertainties, and in some cases, the parallaxes are negative. Therefore, the straightforward approach of inverting the observed parallax is either inaccurate or impossible to an individual star.  Averaging a collection of stars would mitigate the problem that a subset of the stars have negative parallaxes, but this average would be biased due to the large uncertainties. Therefore, a more sophisticated inference is required such as the Bayesian approach described in Section~\ref{sec:Bayes}.

\subsection{Average statistics: a first rough statistical inference}
\label{sec:analyticmodel}
Before using Bayesian inference, we estimate the true mean parallax for several annuli to roughly infer the cluster parallax.


\begin{figure}
\includegraphics[width=0.5\textwidth]{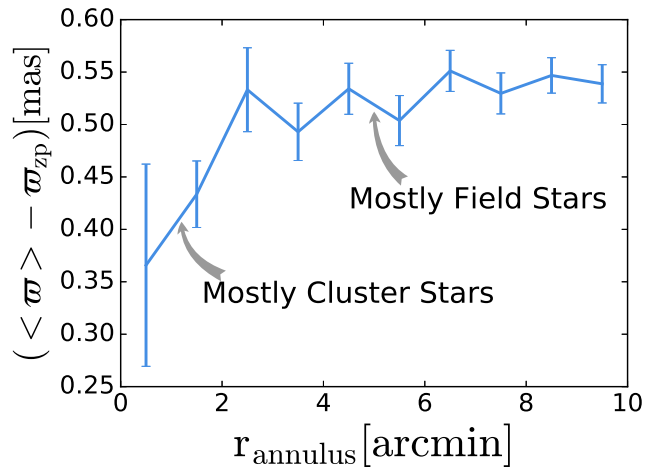}
\caption{Estimates for the true mean parallax for each ring. The mean observed parallax is $\langle \varpi \rangle$, and the parallax zero-point is $\varpi_{\rm zp} = -0.05$ mas. The average parallax for the outer rings most likely represent the field stars, and the inner rings represent the cluster.  Below $\sim$3 arcmin, the average parallax transitions from being dominated by the field stars to the cluster stars. The vertical error bars show the 68\% confidence interval after bootstrapping the average 500 times for each annulus.}
\label{fig:MeanOBSOmega}
\end{figure}

Fig.~\ref{fig:MeanOBSOmega} shows the estimates for the true mean parallax as a function of radius from the cluster centre. After using the selection criteria (Section~\ref{sec:data}), we calculate the mean parallax for each annulus, and adjust it by the parallax zero-point to get an estimate for the true parallax. These averages show a clear indication
that the cluster has a smaller parallax (larger distance) than the
average field star.  The blue line
represents the observed mean parallax $\langle \varpi \rangle$ corrected by the parallax zero-point of $\varpi_{\rm{zp}}=-0.05$ mas for several annuli. For the zero-point, we take the mean of recent estimates \citep{L18,R18,S18,Z18}.  See Section~\ref{sec:Bayes} for more details. Each
 annulus is 1 arcmin in width.  To estimate the uncertainty on
the average, the vertical error bars show the 68\% confidence interval
after bootstrapping the average 500 times for each annulus. If the uncertainties are accurate, then one could just calculate the uncertainty using standard error propagation.  However, as we show below, the uncertainties are inconsistent with the scatter in the data.  Therefore, we choose to use the actual data to report the uncertainties in the average.  For
radii above $\sim$ 3 arcmin, the average parallax seems to be
dominated by the field stars.  Interior to this radius, $\langle
\varpi \rangle$ becomes more influenced by the cluster with decreasing radius.

Fig.~\ref{fig:MeanOBSOmega} also justifies our choice for the size of the inner circle and the outer annulus.  To be clear, the cluster is not limited to a radius of 1 arcmin.  Rather, this is the radius for which we restrict our inference of the cluster parallax.  Based upon Fig.~\ref{fig:MeanOBSOmega}, the cluster clearly extends several arcminutes.  However, our model assumes one density for the cluster.  Therefore, we restrict our inference to the inner region where the density is roughly uniform and consistent with our model assumption.  Conversely, since the cluster clearly extends several arcminutes, we choose the outer ring far enough to be dominated by field stars.  Fig.~\ref{fig:MeanOBSOmega} indicates that 10 arcmin is sufficiently far enough to satisfy this requirement. 

One obvious way to estimate the parallax is using the variance weighted mean. The variance-weighted mean parallax for the inner circle is $0.60\pm0.1$ mas, which is inconsistent with the simple mean. $0.1$ mas represents the uncertainty from bootstrapping, while the standard error propagation gives a smaller uncertainty of 0.01 mas. This suggests that the empirical uncertainty is quite a bit larger than the reported uncertainty (see Section~\ref{sec:discussion} for more details). Given that the variance-weighted mean is compromised by the inaccurate uncertainties, we should use a more complicated method like Bayesian inference to infer the parallax. In the next subsections, we infer the cluster parallax through Bayesian inference.

\subsection{Bayesian analysis}
\label{sec:Bayes}
To infer the Wd1 parallax, the posterior distribution for the model parameters, $\theta$ is the product of the likelihood $\mathcal{L}(data|\theta)$ and a multidimensional prior probability $ P(\theta)$:
\begin{equation}
\label{eq:Bayes}
P(\theta|data) \propto \mathcal{L}(data|\theta)P(\theta)\, .
\end{equation}
Before we fully describe the likelihood model, we define the model parameters and data.  The density of stars (see Fig.~\ref{fig:Ring}) suggests a model for two sets of stars.  One set, the inner circle, contains both cluster and field stars, and the other set, outer ring, includes only field stars.  The outer ring constrains the parameters of the field-star distribution, one of which is the length scale ($L$). $L$ gives an effective length scale for the distribution of field stars, equation~(\ref{eq:lengthscale}).  In practice, this length scale is set by many factors, and one of the main factors is an effective optical depth for extinction along the line of sight \citep{B18}. 

The full set of observations, $data$, includes two data sets: the parallaxes of stars in the inner circle, $\{\varpi_j\}$, and the parallaxes of stars in the outer annulus, $\{\varpi_k\}$. $data$ also includes the number of stars in the inner circle, N$_i$, and the number of stars in the outer annulus, N$_\text{o}$. $\{\sigma_{\varpi_j}\}$ and $\{\sigma_{\varpi_k}\}$ are the parallax uncertainties for the inner region and the outer annulus. We consider the parallax uncertainties to be
fixed parameters and the dependencies on the uncertainties are omitted in the following equations for brevity. The model parameters, $\theta$, are the cluster parallax, $\varpi_{\text{cl}}$ (mas), density of the cluster stars in the inner circle, $n_{\text{cl}}$ (number per square arcminute), density of the field stars in the outer ring, $n_{\text{f}}$ (number per square arcminute), which we assume to be similar in the inner ring, the parallax zero-point of the cluster, $\varpi_{\text{zp}}$ (mas), the field-star length scale, $L$ (kpc), and the field-star offset, $\varpi_{\text{os}}$ (mas). The field-star offset ($\varpi_{\text{os}}$) includes the parallax zero-point but it also includes other possible systematics that affects the field-star distribution but not the cluster parallax; we elaborate more on this later. Naturally, $n_{\text{cl}}$ is a function of radius from the centre of the cluster.  Rather than assuming a radial profile, we consider annuli and infer an average number density for each annulus.  Specifically, henceforth, $n_{\text{cl}}$ refers to the average density of the central arcminute of the cluster. Each set also has a nuisance parameter. The nuisance parameters, $\eta$, are the set of true parallaxes for the inner circle, $\{\hat{\varpi}_j\}$ (mas), and the set of true parallaxes for the outer annulus, $\{\hat{\varpi}_k\}$ (mas).

\begin{figure*}
\includegraphics[width=6.5in]{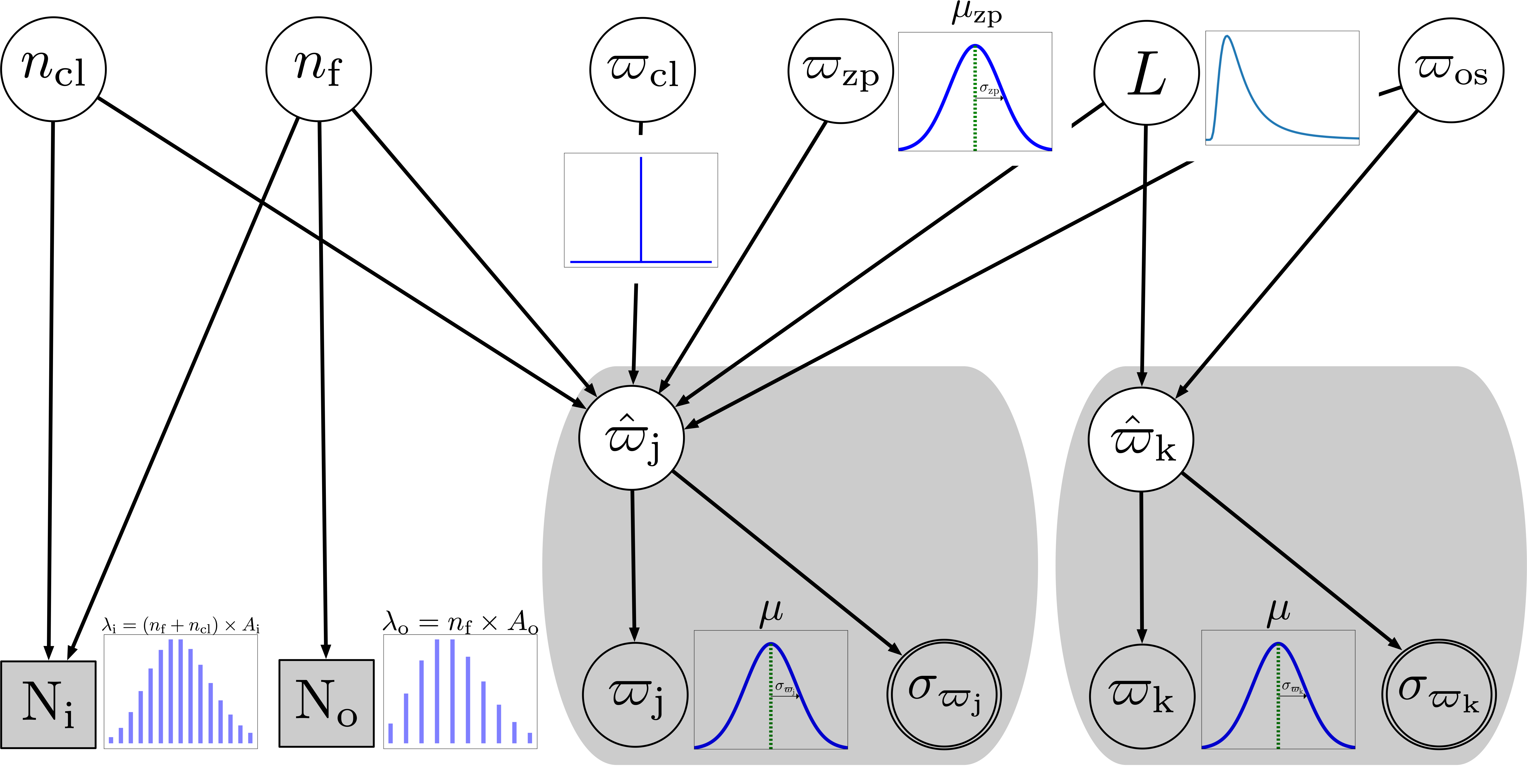}
\caption{Probabilistic graphical model for the Bayesian model.  Arrows show the dependence of variables, circles indicate continuous variables, and double circles show fixed values.  The top row shows the model parameters; for the first six (the cluster parallax, $\varpi_{\text{cl}}$, density of the cluster stars in the inner circle, $n_{\text{cl}}$, density of the field stars, $n_{\text{f}}$, the parallax zero-point of the cluster, $\varpi_{\text{zp}}$, the field-star length scale, $L$, and the field-star offset, $\varpi_{\text{os}}$) we infer a posterior distribution. The middle row shows the latent, or nuisance, parameters (the set of true parallaxes for the inner circle, $\{\hat{\varpi}_j\}$, and the set of true parallaxes for the outer annulus, $\{\hat{\varpi}_k\}$).  The bottom row shows the observations (the number of stars in the inner circle, N$_\text{i}$, the number of stars in the outer annulus, N$_\text{o}$, the parallaxes for the inner circle, $\{\varpi_j\}$, and the parallaxes for the outer annulus, $\{\varpi_k\}$). This diagram maps out the dependencies when deriving the conditional probabilities.}
\label{fig:PGM}
\end{figure*}

The probabilistic graphical model (PGM), Fig.~\ref{fig:PGM}, shows the interdependence among the observations, the model parameters,  and the nuisance parameters. The likelihood probability is:
\begin{equation}
\label{eq:likelihood}
\mathcal{L}(data|\theta)=\mathcal{L}({\rm N}_\text{i}|n_{\rm cl},n_{\rm f})\mathcal{L}({\rm N_o}|n_{\rm f})\mathcal{L}(\{\varpi_{j}\}|\theta)\mathcal{L}(\{\varpi_{k}\}|\theta)\, .
\raisetag{1.2\normalbaselineskip}
\end{equation}
Each component of the likelihood on the RHS is a likelihood for a particular set of data.  The parameters after $|$ represent the set of model parameters which determine the data in the model.  If the data depend upon the full set of model parameters, then $\theta$ appears, otherwise we include only the model parameters that matter for each likelihood
component.  For example, the number of stars in the inner circle, N$_\text{i}$, depends only upon the cluster density, $n_{\rm cl}$, and field density, $n_{\rm f}$.

To describe the model for the observations, first, consider the bottom row in Fig.~\ref{fig:PGM}; it shows all observed parameters.  Near the centre is $\{ \varpi_j\}$, which represents the data for inner circle.  Each observed parallax is drawn from a Gaussian distribution centred on the true parallax; each star has its own true parallax, $\hat{\varpi}_j$; therefore, there is a set of many true parallaxes for the inner region, 
$\{ \hat{\varpi}_j\}$.  Since the central region has both field and cluster stars, each true parallax is either drawn from the cluster or the field-star parent distributions.  A priori, it is not clear which star is associated with the cluster or field-star distributions.  However, the fraction of stars associated with the cluster is $n_{\rm cl}/(n_{\rm cl} + n_{\rm f})$ and the fraction associated with the field stars is $n_{\rm f}/(n_{\rm cl} + n_{\rm f})$.  Hence, the density of cluster and field stars are important parameters in the generation of the data.  Unfortunately, modelling just the inner circle does not constrain the four model parameters along the top row.

Modelling the outer ring provides constraints on the field-star parameters.  This then, in combination with the inner circle data, provides a unique constraint on the cluster parameters.  Each observation of the outer annulus, $\{ \varpi_k\}$, is also drawn from a Gaussian, and the true field-star parallax is drawn from the field-star distribution.  Our assumption is that all stars in the outer ring are drawn from the field star distribution.  The number of stars in the outer ring is simply given by the field-star density, while the inner circle is a weighted combination of the cluster and field-star densities.  Hence, to constrain the cluster density and ultimately the fraction of cluster stars in the inner circle, the likelihood must model both the number of stars in the inner circle and the outer ring.

The first two likelihoods in equation~(\ref{eq:likelihood}), $\mathcal{L}({\rm N}_\text{i}|n_{\rm cl},n_{\rm f})$ and $\mathcal{L}({\rm N_o}|n_{\rm f})$, represent the number of stars in the inner circle and the outer ring:
 \begin{equation}
\mathcal{L}(N_\text{i}|n_{\rm cl},n_{\rm f})=\frac{\lambda_
{\text{i}}^{N_\text{i}}e^{-\lambda_\text{i}}}{N_\text{i}!}\, ,
\end{equation}
and 
\begin{equation}
\mathcal{L}(N_o|n_{\rm f})=\frac{\lambda_
{\text{o}}^{N_o}e^{-\lambda_\text{o}}}{N_o!}\, ,
\end{equation}
where the expected number of stars in the inner circle is $\lambda_\text{i}=(n_{\rm f}+n_{\rm cl})\times {\rm A}_\text{i}$ and the expected number of stars in the outer ring is $\lambda_\text{o}=n_{\rm f} \times {\rm A}_\text{o}$. A$_\text{i}$ is the area of the inner circle and  A$_\text{o}$ is the area of the outer ring. $\mathcal{L}(\{\varpi_{\rm{k}}\}|\theta$) in equation~(\ref{eq:likelihood}) is the likelihood for the outer set of data:
\begin{equation}
\label{eq:OuterLikelihoodtot}
\mathcal{L}(\{\varpi_k\}|\theta)=
\prod_k \mathcal{L}_{k}(\varpi_k|\theta)\, .
\end{equation}
PGM provides a map of how to further deconstruct the likelihood using the conditional probability theorem:
\begin{equation}
\label{eq:OuterLikelihood}
\mathcal{L}_{k}(\varpi_k|\theta)=\int P_k(\varpi_k|\hat{\varpi}_k) \times P_k(\hat{\varpi}_k|\theta) d\hat{\varpi}_{k}
 \, .
\end{equation}

The first term in equation~(\ref{eq:OuterLikelihood}), $P_k(\varpi_k|\hat{\varpi}_k)$, is the probability of observing any parallax for the $k$th star in the outer ring:
\begin{equation}
\label{eq:OuterData}
P_k(\varpi_k|\hat{\varpi}_k)= 
\frac{1}{\sqrt{2 \pi }{\sigma_{\varpi_k}}} \exp \left [ \frac{-(\varpi_k -\hat{\varpi}_k)^2}{2 {\sigma^2_{\varpi_k}}} \right ] \, ,
\end{equation}
where $\sigma_{\varpi_k}$ is the parallax uncertainties for the outer annulus. In this work, we consider the parallax uncertainties to be fixed parameters. The second term in equation~\ref{eq:OuterLikelihood}, $P(\hat{\varpi}_k|\theta)$, represents the field star distribution.  The PGM shows that this distribution only depends upon two model parameters, $L$ and $\varpi_{\text{os}}$; hence, $P(\hat{\varpi}_k|L,\varpi_{\text{os}})$. If one considers an image populated with stars, then the total number of stars in the image is given by $N = {\rm FOV} \int n r^2$dr, where FOV is the field of view in square radians, $n$ is the number density of stars, and r is the distance from the sun.  If $n$ is constant, then any random star in the image is drawn from a probability distribution of $P(r) \propto r^2$. This distribution is assumed to fall off exponentially, $\exp({-r/L})$, due to various effects including the {\it Gaia} selection function, attenuation due to dust or a combination of both. Therefore, the distribution of the field stars is
\begin{equation}
\label{eq:lengthscale}
P(r|L)=\frac{1}{2L^3} r^2  \exp(-r/L)\, .
\end{equation}
After transforming from distance to parallax, the field-star distribution becomes
\begin{equation}
\label{eq:OuterField}
\begin{split}
P(\hat{\varpi}_k|L,\varpi_{\text{os}})=\frac{1}{2L^3} \frac{\exp[-1/((\hat{\varpi}_k-\varpi_{\text{os}}) L)]}{\hat{\varpi}_k^4}\, .
\end{split}
\end{equation}
Together, equations~(\ref{eq:OuterData}) and (\ref{eq:OuterField}) represent the likelihood for the outer ring, given in equation~(\ref{eq:OuterLikelihood}). 
The zero-point in equation~(\ref{eq:OuterField}), $\varpi_{\text{os}}$ represents an offset for the field star distribution, and it may or may not be the same as the zero-point, $\varpi_{\text{zp}}$, for the cluster members.  $\varpi_{\text{zp}}$ represents a zero-point associated with instrumental and analysis biases \citep{L18}.  The parallax distribution for field stars, equation~(\ref{eq:OuterField}), assumes that the distribution approaches zero at $\hat{\varpi}_k = 0$; however, it may not do so for several reasons.  Certainly, one part is due to the same instrumental and analysis biases that impact the cluster members.  In addition, sightlines through the plane of the Galaxy likely have a distribution of field stars that is more complicated than equation~(\ref{eq:OuterField}).  For example, while uniformly distributed dust may cause an exponential attenuation with one scale, $L$, a very dusty star-forming region in a spiral arm will likely present a wall, beyond which we cannot see any stars.  This would manifest as an abrupt truncation of the field-star distribution at some finite, positive parallax.  For this reason, the zero-points for the field stars and the cluster must remain separate variables.

To find the likelihood distribution for the outer ring we marginalize over the nuisance parameters, $\eta$. In the PGM, Fig.~\ref{fig:PGM}, the nuisance parameters are the true parallaxes, $\{\hat{\varpi}_k\}$. 
The convolution of the Gaussian and the true field distribution is not analytic and requires a numerical solution.

To derive the likelihood for the inner ring,  $\mathcal{L}(\{\varpi_{j}\}|\theta$) in equation~(\ref{eq:likelihood}), we use equations~(\ref{eq:OuterLikelihoodtot}) and (\ref{eq:OuterLikelihood}), but with a change of index from $k$ to $j$. The probability of observing any parallax for the $j$th star in the inner circle $P_k(\varpi_k|\hat{\varpi}_k)$ is also same as equation~(\ref{eq:OuterData}) with a simple exchange of index from $k$ to $j$.

Next, we propose a distribution for the true parallaxes of the inner circle given the model parameters, $P_j(\hat{\varpi}_j|\theta)$.
Once again, the inner circle is composed of cluster and field stars. 
A star in the inner circle is either drawn from the cluster or from the field-star distributions, and the weighting for each draw is proportional to the density of the respective population. Hence, the distribution for the true parallaxes is
\begin{multline}
\label{eq:inner}
P_j(\hat{\varpi}_j|\theta)=
\left(\frac{n_{\rm cl}}{n_{\rm cl}+n_{\rm f}}\right) P(\hat{\varpi}_j|\varpi_{\rm cl},\varpi_{\text{zp}})\\
+ \left(\frac{n_{\rm f}}{n_{\rm cl}+n_{\rm f}}\right) P(\hat{\varpi}_j|L,\varpi_{\rm{os}}) \, .
\end{multline}
Since the size of the cluster is much smaller than the distance to the cluster, we assume a delta function for the cluster true-parallax distribution at $\varpi_{\rm cl}$; $P(\hat{\varpi}_j|\varpi_{\rm cl},\varpi_{\text{zp}})= \delta(\hat{\varpi}_j-\varpi_{\rm cl}-\varpi_{\text{zp}})$. The second term represents the portion associated with the field stars. To find the likelihood distribution for the inner ring, we marginalize over the nuisance parameter for the inner ring, $\{\hat{\varpi}_j\}$. The convolution of a Gaussian and a delta function is analytical and a convolution of a Gaussian and the field-star distribution requires a numerical integration.

To find the posterior distribution, equations~(\ref{eq:Bayes}), we choose uniform positive prior distributions for $n_{\rm cl}$, $n_{\rm f}$, $\varpi_{\rm cl}$, $\varpi_{\rm os}$, and $L$. \citet{L18} found that the zero-point is a function of colour, magnitude and position;  hence, the  zero-point  has significant variance. 
However, there are no reference objects in our field to find the zero-point.  Therefore, we must use prior information to estimate the zero-point of the cluster.  For simplicity, we assume that the zero-point distribution for DR2 fields is Gaussian:
\begin{equation}
\label{eq:ZeroPointPrior}
P(\varpi_{\text{zp}}|\mu_{\text{zp}}) =
\frac{1}{\sqrt{2 \pi} \sigma_{\text{zp}}} \exp \left [\frac{-(\varpi_{\text{zp}} -
  \mu_{\text{zp}})^2}{2\sigma_{\text{zp}}^2} \right ] \, , 
\end{equation}
where $\mu_{\text{zp}}$ and $\sigma_{\text{zp}}$ are the mean and variance for the parallax zero-point. The two most effective means to calculate zero-point are to either use background quasars or to use independent distance measurements.  Wd1 is in the Galactic plane, so there are no background quasars, and the current distance estimates for Wd1 are too uncertain to use to constrain the zero-point.  Therefore, we will use previous analyses to estimate the zero-point distribution. \citet{L18} used quasars to infer the
zero-point. They found an average of $-$0.029 mas. \citet{R18} inferred a zero-point of $-$0.046 $\pm$ 0.013 mas for the Cepheid sample. \citet{Z18} compared the distances inferred from astroseismology
to infer a zero-point of $-$0.0528 $\pm$ 0.0024
mas. \citet{S18} also reported the zero-point of $-$0.082 $\pm$ 0.033 mas from eclipsing binaries. The mean of the above four
investigations is $\mu_{\text{zp}}$ = $-$0.05
mas, and \citet{L18} found that the variation for the zero-point across many fields is $\sigma_{\text{zp}} = 0.043$ mas \citep{L18}. The data that we use in our inference cannot constrain the zero-point.  Therefore, when we infer the posterior distribution for $\varpi_{\text{zp}}$, it will merely reflect this prior distribution.

\subsection{Numerical solution for the posterior distribution}
\label{sec:NumericalS}
To find the posterior distribution, we use a six dimensional Monte Carlo Markov chain package (MCMC), \textsc{emcee} \citep{G10,F13} to infer six model parameters ($n_{\rm cl},n_{\rm f}$, $\varpi_{\rm cl}$, $\varpi_{{\rm zp}}$, $L$, and $\varpi_{\text{os}}$). For each step in the chain, \textsc{emcee} evaluates the posterior by calculating the likelihood for the inner circle and the likelihood for the outer ring. Both likelihoods require the convolution of a Gaussian with the true field distribution. Evaluating these integrals is time intensive. Instead of calculating the integrals at every step in the chain, we create look-up tables for each integral.

Each object has its own look-up table evaluated at a grid of points in $L$.  For each trial of $L$ in the MCMC, we find the convolution by first-order interpolation in the look-up table.  To construct each look-up table, we use trapezoid numerical integration, which requires bounds of integration.  Formally, the bounds extend from $\hat{\varpi} = -\infty$ to $\hat{\varpi} = \infty$, but that is not practical for trapezoid numerical integration.  Fortunately, the integrands in the likelihoods have a peak and fall off quickly on either side of this peak.  To ensure that the numerical integration adequately samples this peak, we set the bounds of integration to be centred on the peak and have a width that extends just outside the peak.  To roughly estimate the position of the peak and the extent of the bounds, we approximate the integrand as the convolution of two Gaussians.  The mode and width of the first Gaussian is straightforward, $\mu_1 = \varpi$ and $\sigma_1 = \sigma_{\varpi}$.  The mode of the field star distribution is $\mu_2 =0.25/L$, and the width is $\sigma_2=0.5/L$.  In the two Gaussian approximation, the mode of the integrand is roughly at $\mu = p\mu_1+q\mu_2$ and the width is roughly $\sigma = \sqrt{p\sigma_1^2+q\sigma_2^2+p\mu_1^2+q\mu_2^2-\mu^2}$ where weighting factors are $p=\frac{4(L\sigma)^2}{1+4(L \sigma)^2} $ and $q=\frac{1}{1+4(L\sigma)^2}$. Therefore, we integrate from $\hat{\varpi} = \mu - 2\sigma$ to $\hat{\varpi} = \mu +2\sigma$.

For each MCMC run, we use 100 walkers, 2500 steps each, and we burn 500 of those. For the results presented in Section~\ref{sec:results}, the acceptance fraction is in the $\alpha=0.5$ range.

\section{Parallax and Distance to Westerlund 1}\label{sec:results}

\begin{figure*}
\includegraphics[width=6.5in]{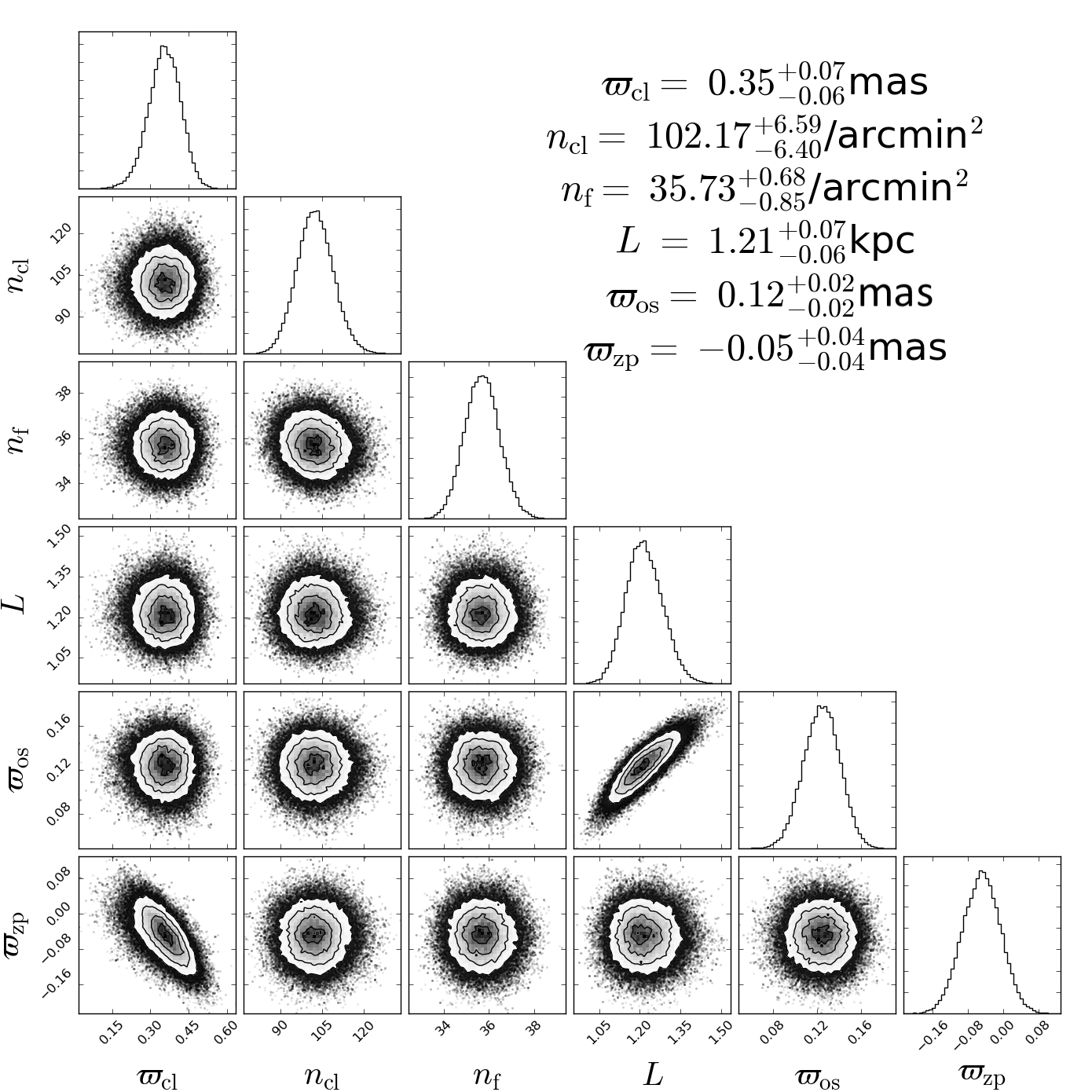}
\caption{Posterior distribution for the six-parameter model. We report the mode and the highest density 68\% confidence interval for the cluster parallax ($\varpi_{\text{cl}}$), the cluster density ($n_{\text{cl}}$), the field-star density ($n_\text{f}$), the field-star length scale ($L$), the field-star offset ($\varpi_{\text{os}}$), and the parallax zero-point of the cluster ($\varpi_{\text{zp}}$). The parallax of the cluster is $\varpi_{\text{cl}}=0.35^{+0.07}_{-0.06}$ mas, which corresponds to a distance of $R=2.6^{+0.6}_{-0.4}$ kpc. The posterior for the cluster zero-point, $\varpi_{\text{zp}}$, reflects the prior; there is no information in this data to constrain this parameter.}
\label{fig:Corner}
\end{figure*}

Fig.~\ref{fig:Corner} shows the posterior distribution for $\varpi_{\text{cl}}$, $n_{\text{cl}}$, $n_\text{f}$, $L$, $\varpi_{\text{os}}$, and $\varpi_{\text{zp}}$. The two regions used to constrain these parameters are an inner circle  centred on the position of Wd1 and with a radius of 1 arcmin, and an outer annulus from 9 to 10 arcmin.  The values in the top right corner show the mode and the highest 68\% density interval (HDI) for marginalized distributions. The parallax of the cluster is $\varpi_{\text{cl}}=0.35^{+0.07}_{-0.06}$ mas, which corresponds to a distance of $2.6^{+0.6}_{-0.4}$ kpc, density of the cluster is $n_{\text{cl}}=102.17^{+6.59}_{-6.40}$ stars per square arcminute, density of field stars is $n_\text{f}=35.73^{+0.68}_{-0.85}$ stars per square arcminute, the parallax zero-point of the cluster is $\varpi_{\text{zp}}=-0.05\pm{0.04}$ mas, the field-star length scale is $L=1.21\pm{0.02}$ kpc, and the field-star offset is $\varpi_{\rm{os}}=0.12\pm{0.02}$ mas. The posterior shows a single-peaked marginalized probability distribution for all parameters. 

\begin{figure}
\includegraphics[width=0.5\textwidth]{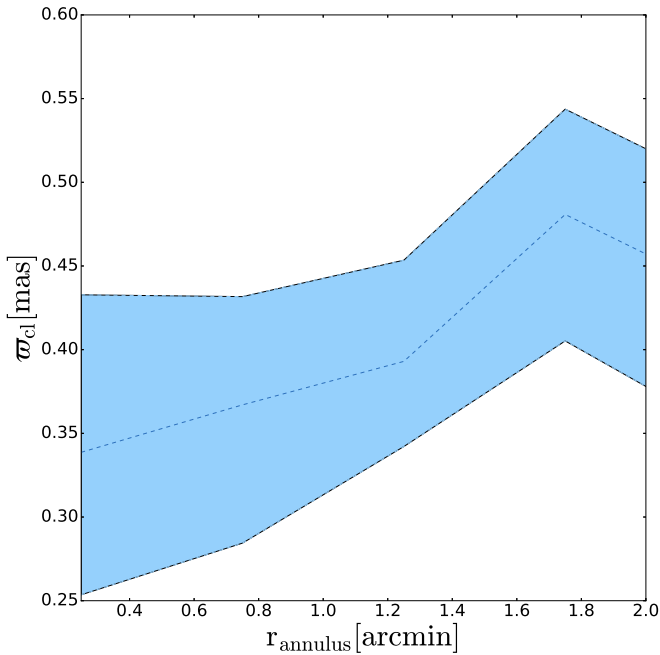}
\caption{Bayesian inferred cluster parallax for each ring. Top and bottom dashed lines represent the highest density 68\% confidence interval. All rings bellow 1.5 arcmin are consistent with the inference from stars within 1 arcmin.}
\label{fig:omegaclBays}
\end{figure}

Presumably, the cluster density is a function of radius.  Since we model the cluster density with one average density and not a radial profile, there is a potential for bias to affect the inference.  As long as the width of each annulus is small compared to the change in density, then approximating the density in each ring with an average annulus should work well.  To test this hypothesis, we infer the full posterior distribution for several inner annuli.  For each inference, we use the outer annulus from 9 to 10 arcmin to constrain the field-star parameters.  Fig.~\ref{fig:omegaclBays} shows the inferred parallaxes for each annulus. The rings extend from 0 to 0.5, 0.5 to 1.0, 1.0 to 1.5, and 1.5 to 2.0 arcmin with 8, 34, 105, and 141 total number of stars, respectively.  In terms of parallax, the $68\%$ highest density intervals are [0.25,  0.43], [0.28, 0.43], [0.34, 0.45], and [0.40, 0.54] mas.  All rings below 1.5 arcmin are consistent with our main result of 0.35$^{+0.07}_{-0.06}$ mas from using an inner circle with radius 1 arcmin (Fig.~\ref{fig:Corner}), and rings above 1.5 arcmin more likely represent the field stars parallax. Therefore, we conclude that modelling the average cluster density rather than a radial density profile is sufficient for the chosen inner region sizes.

\begin{figure*}
\includegraphics[width=6.5in]{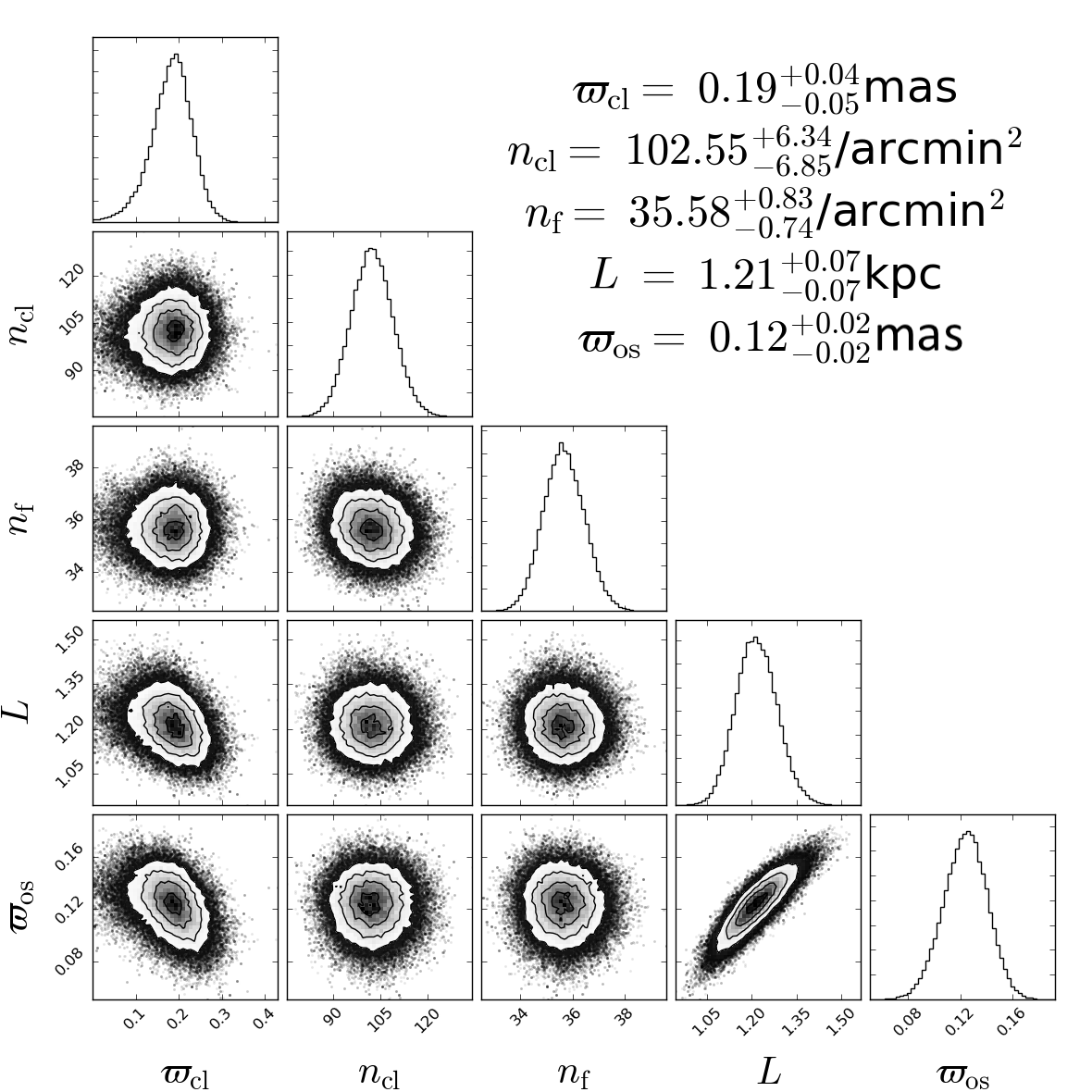}
\caption{Same as Fig.~\ref{fig:Corner}, but we assume that the offset for the field-star distribution, $\varpi_{\text{os}}$ is equal to the instrumental zero-point, $\varpi_{\text{zp}}$. With this assumption, the cluster parallax is $\varpi_{\text{cl}}=0.19^{+0.04}_{-0.05}$ mas, which corresponds to distance of $R=4.9^{+1.5}_{-1.0}$ kpc. However, the field-star distribution has systematics other than the instrumental systematic; therefore, Fig.~\ref{fig:Corner} represents the most likely inference.}
\label{fig:Corner2}
\end{figure*}

\section{Discussion}\label{sec:discussion}
The results in Section~\ref{sec:results} have significant implications for both the distance to Wd1 and the distribution of field stars. The inferred $L$ is $1.21^{+0.07}_{-0.06}$ kpc, which is consistent with the model of \citet{B18}, $L_{\text{BJ}} =$1.38 kpc.  The 2$\sigma$ difference is not that surprising given that the \citet{B18} estimate for $L$ varies as a function of Galactic longitude and latitude (l,b), and it does not include clusters. Also, given that the \citet{B18} length scale was derived from a model of the Galaxy before the era of accurate {\it Gaia} parallaxes, it is encouraging that the \citet{B18} model is only 2$\sigma$ away from the inferred value. Using \citet{B18} length scale, $L$=1.38 kpc, and using the same parallax zero-point for the whole region, we find that the cluster parallax is $\sim0.39\pm{0.05}\pm{0.04}$(sys) mas, which differs by only 1$\sigma$ from the value when we infer $L$ as well.  While the results are formally consistent, using the \citet{B18} length scale, skews the cluster parallax towards the average field-star distribution.

Our inferred parallax of the cluster is $\varpi_{\text{cl}}=0.35^{+0.07}_{-0.06}$ mas, which corresponds to a distance of $2.6^{+0.6}_{-0.4}$ kpc. For this inference, we used Bayesian inference with a six-parameter model; there are other, simpler statistical inferences, but these typically ignore known contamination and systematics.  These simpler inferences are useful in checking the results of our Bayesian inference.  For example, we calculate the average distance for all stars within 1 arcmin from the cluster centre, where the individual distances comes from \citet{B18}.  To estimate the uncertainty, we bootstrap the average.  In this way, the average distance for these stars is $3.2\pm{0.06}(\rm{stat})\pm{0.4}(\rm{sys})$ kpc; 25\% of our posterior distribution contains distances larger than this simple mean. The systematic uncertainty is due to the zero-point uncertainty \citep{L18} and it  dominates the uncertainty. Or another approach is to calculate the true mean parallax for the inner circle with 1 arcmin radius (see the first point in Fig.~\ref{fig:MeanOBSOmega}).  This gives $0.36\pm{0.09}\text{(stat)}\pm{0.04}(\text{sys})$ mas; this simple mean is within the 68\% confidence interval of our inferred parallax. Once again, the systematic uncertainty is due to the uncertainty in the zero-point. While these simple approaches provide a good check on our result, these naive calculations are not sufficient.  For one, many of the parallaxes are negative.  Secondly, we infer that at least 1/4 of the stars in the inner arcminute are field stars.  Therefore, to infer the distance to the cluster one must use Bayesian inference and model both the field stars and the cluster stars.

Wd1 is located at a Galactic longitude of $\ell=339.55\degr$. Given this longitude and the new inferred distance of $2.6^{+0.6}_{-0.4}$ kpc ($\sim$ 8500 ly), Wd1 most likely lies in the Scutum-Centaurus arm, which is one of the major spiral arms of the Milky Way \citep{b14,U14, V14}. This may be an independent way to confirm the new inferred distance of $2.6^{+0.6}_{-0.4}$ kpc to Wd1.

Based upon the prior distribution, the parallax zero-point for the cluster is $\varpi_{\rm{zp}}=-0.05\pm{0.04}$ mas; in contrast, we infer a field-star offset of $\varpi_{\rm{os}}=0.12\pm{0.02}$ mas. In our model, $\varpi_{\rm{os}}$ represents a combination of the instrumental zero-point, $\varpi_{\text{zp}}$, and a truncation of the field-star distribution. It could represent the zero-point for the entire region.  However, 0.12 mas is much larger than the average of all other previous analyses (mean of $-0.05$ mas).  Therefore, the most likely conclusion is that $\varpi_{\rm{os}}$ is dominated by a truncation of the field star distribution.  This would occur if the line of sight toward this region intersects an exceptionally dense region of dust at a true parallax around 0.17 mas.  This scenario is consistent with the fact that this line of sight is in the plane of the Galaxy. Alternatively, if we force $\varpi_{\text{zp}} = \varpi_{\text{os}}$, then we find that the parallax of the cluster is $\varpi_{\rm{cl}}=0.19^{+0.04}_{-0.05}$ mas, which corresponds to a distance of $R=4.9^{+1.5}_{-1.0}$ kpc (see Fig.~\ref{fig:Corner2}). Again, the most likely scenario is that $\varpi_{\text{os}} \neq \varpi_{\text{zp}}$; therefore, the most likely cluster parallax is $\varpi_{\text{zp}} = 0.35^{+0.07}_{-0.06}$ mas.

\begin{figure}
\includegraphics[width=0.5\textwidth]{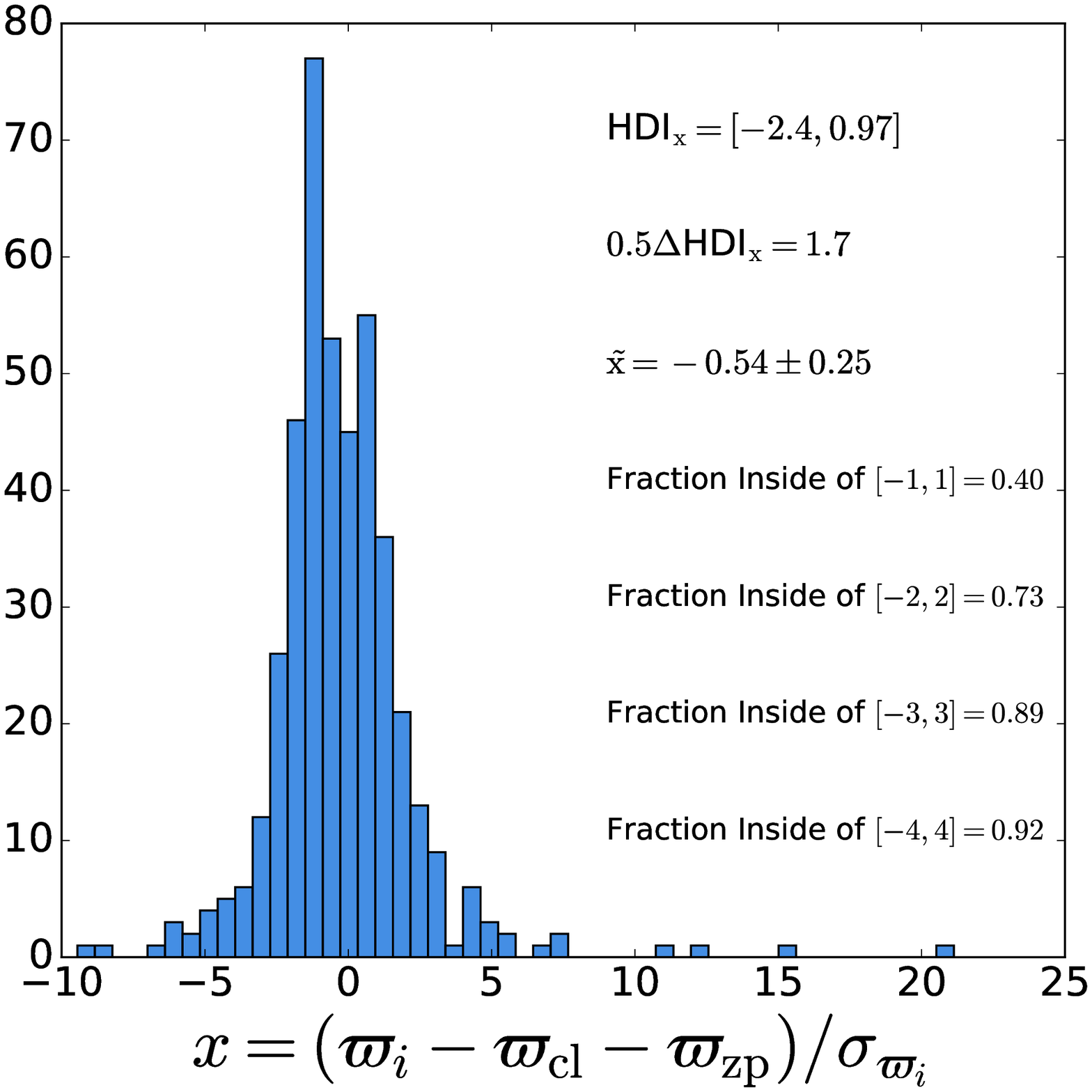}
\caption{Histogram of scale factor $x$ for all sources within 1 arcmin of centre of the cluster. The empirical uncertainty for stars in Wd1 is 1.7 times larger than the DR2 solution. If uncertainties represents the full variation in the data, one would expect the distribution to have a 68\% HDI that ranges from -1 to 1. However, the variances are significantly smaller than expected variation for all of the stars within 1 arcmin of Wd1.}
\label{fig:Xscale}
\end{figure}

$\sigma_\varpi$ is the expected parallax uncertainty, but the empirical uncertainty is different. For example, \citet{L18} found that the empirical uncertainty is 1.081 times larger than the expected value for quasars in DR2. Similarly, we measure the empirical uncertainty distribution for all stars within 1 arcmin of Wd1. Fig.~\ref{fig:Xscale} shows the parallaxes centred on the cluster and weighted by the uncertainty:
\begin{equation}
x = \frac{\varpi_\text{i} - \varpi_{\rm cl} - \varpi_{\rm zp}}{\sigma_{\varpi_\text{i}}} \, .
\end{equation}
If the uncertainties are accurate, uncorrelated and random, then the distribution of $x$ should be Gaussian with $\sigma_x = 1$.  In other words, 68\% of $x$ should be within the interval $[-1,1]$.  Fig.~\ref{fig:Xscale} reports the percent of the distribution inside of 1, 2, 3, and 4$\sigma$.  If uncertainties represent the full variation and the distribution is Gaussian, then one would expect the distribution to have 68\% of the data inside 1$\sigma$, 95\% inside 2$\sigma$, etc. However, the probabilities are smaller than one would expect.  In fact, we calculate that the 68\% highest density interval for $x$, HDI$_x$ is $[-2.4,0.97]$.  The half width of this interval is 1.7.  In other words, the empirical uncertainties are 1.7 times larger than the expected uncertainties.  This suggests that there are significant problems with the uncertainty estimate in the region of Wd1.  At the moment, it is not clear what is causing this inaccuracy.  The high extinction (red colours), crowding, and binarity may be three important factors.  Without having the raw data, it is difficult to investigate this problem further, so we proceed with our inference, keeping in mind that the DR2 uncertainties are likely too small by a factor of 1.7.

The fact that the empirical uncertainties are 1.7 larger than the calculated uncertainties suggests that there is a problem with the five-parameter astrometric model for these stars.  This could be due to any number of issues, including the excess noise model, binarity, or the degrees of freedom bug \citep{L18}.  Formally, the astrometric excess noise is a way to incorporate extra empirical noise.  The DR2 pipeline uses error propagation to include the excess noise into $\sigma_{\varpi_\text{i}}$ of source i.  However, even after the excess noise is included we find that there is still significant excess empirical noise.  This discrepancy may be due to the specific model assumed for the excess noise in the DR2 pipeline.  However, equation~120 of \citet{L12} assumes that the excess noise for each observation is uncorrelated.  Hence, more observations will reduce the uncertainty by the square root of the observations.  However, if the true excess noise is correlated among the observations, more data will not reduce the uncertainty; the excess noise would represent a floor on the uncertainty.  The assumption that the excess noise for each observation is uncorrelated may be why the empirical uncertainties for Wd1 are 1.7 times larger than the calculated uncertainties.

The inferred distance to Wd1 is $2.6^{+0.6}_{-0.4}$ kpc, which represents the highest precision distance estimate for  Wd1 that has been published. Historical estimates to Wd1 range from 1.0 to 5.5 kpc (see Section~\ref{sec:intro}). Recently, \citet{C05} estimated that the distance to Wd1 ranged from 2 to 5.5 kpc. The bounds given by \citet{C05} corresponds to a factor  of $\sim$2.8 in distance; in contrast, the precision in the {\it Gaia} DR2 inferred distance is 18\%. One can use the new distance to infer the fundamental parameters of the cluster such as luminosity, mass, and age via isochrone fitting. In this manuscript, we do not perform isochrone fitting.  Instead, we estimate these fundamental parameters using two techniques.  First, we scale previous estimates using the new distance, and we infer the luminosity, mass, and age of two bright stars in Wd1.  The estimates of two bright stars provides a good proxy for the whole cluster. 

The 18\% precision in distance will lead to $\sigma_L/L \approx 36\%$ precision in luminosity. For stars below about 20 M$_\odot$, the main-sequence luminosity for a given mass should scale as $L \propto M^{3.5}$. Therefore, the corresponding uncertainties in mass and age are $\sigma_M/M \approx 10.3\%$ and $\sigma_t/t \approx 3.7\%$, respectively. For stars above about 55 M$_\odot$, $L \propto M$, and in this case, the range of mass estimates has the same precision as the luminosity. For the highest masses, the age depends very weakly on mass or luminosity because these stars have very similar lifetimes around 3 Myr.

The new {\it Gaia} DR2 distance also provides strong constraints on the luminosity, mass, and age of cluster members. Wd1 hosts a diverse population of evolved massive stars such as WR stars, red and blue supergiants, YHGs, an LBV, and a magnetar.  Previous studies have inferred a turn-off mass of around 40 M$_\odot$ and cluster age of 3.5-5 Myr with a presumed distance of around 5 kpc \citep{C05,C06,R09,NC10,L13}.  These estimates were based on modelling the luminosity and temperatures of YHGs, RSGs, and WR stars. By association, this would imply that the magnetar progenitor had an initial mass of $>$40 M$_\odot$ \citep{M06,R10}. On the other hand, \citet{K12} found a progenitor mass of $\sim$ 40 M$_\odot$ and a distance value of 3.7$\pm$0.6 kpc by studying eclipsing binaries in the Wd1.

Without re-evaluating the bolometric and extinction corrections for each star, the new distance of 2.6$^{+0.6}_{-0.4}$ kpc reduces all luminosities by -0.58 dex as compared to a distance of 5 kpc. Using single-star stellar evolution models \citep{Bd11}, we now infer the mass, age, and corresponding main-sequence turn-off mass for two of the brightest stars in Wd1, the LBV W243 and a YHG4.  The inferred log$(L/L_{\rm{M_\odot}})$ for W243 is 5.2$\pm{0.1}$ and for YHG4 is 5.4$\pm{0.1}$.  The spectral type of W243 is B2I (to A2I) \citep{W87,CN04} and for YHG4 is F2Ia$^{+}$.  This corresponds to temperatures of 9.17 kK (to 17.58 kK) and 7.2 kK, respectively.  For these temperatures, using single-star stellar evolution models \citep{Bd11}, the masses are 23.9$^{+2.8}_{-3.2}\rm{M_\odot}$ for W243 and 28.6$^{+4.1}_{-4.8}\rm{M_\odot}$ for YHG4, the corresponding ages are 7.6$^{+1.0}_{-0.9}$ and 5.5$^{+1.6}_{-0.5}$ Myr, respectively.  While we did not fit isochrones, these ages would correspond to isochrones with main-sequence turn-off masses of 22.3$^{+2.2}_{-2.4}\rm{M_\odot}$ (W243) and 25.9$^{+2.8}_{-3.2}\rm{M_\odot}$ (YHG4).

\begin{figure*}
\includegraphics[width=6.5in]{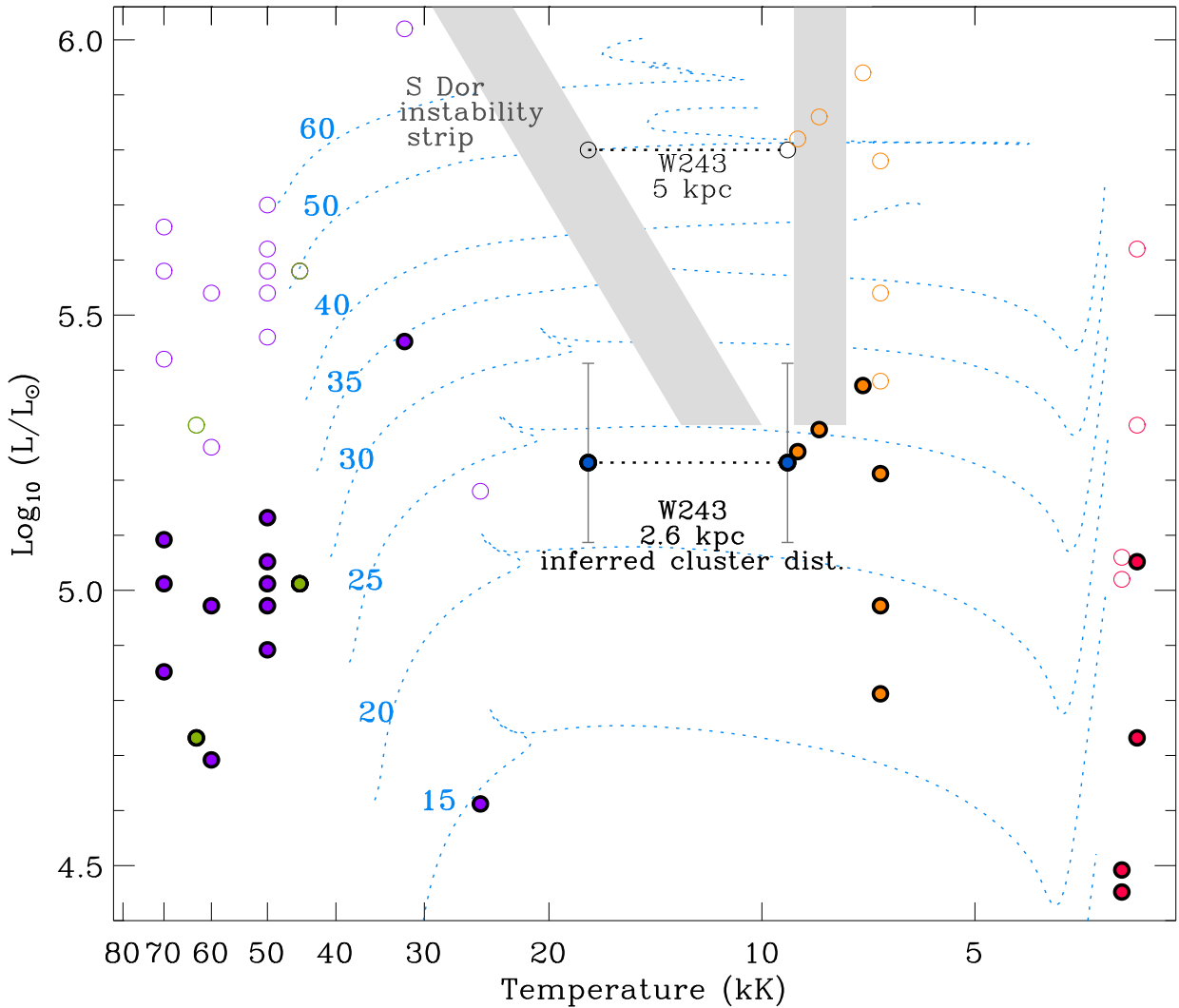}
\caption{The HR diagram for evolved stars in Westerlund 1, including the LBV, W243.  The open circles show the luminosities when $d = 5$ kpc, and the filled circles show the luminosities with the new {\it Gaia} DR2 distance of 2.6$^{+0.6}_{-0.4}$ kpc.  All filled circles have the same uncertainty as W243. The orange symbols represent YHGs, the red symbols represent RSGs, the purple symbols represent WNs, and the green circles represent WCs \citep{C06,F18}. The grey boxes show the locations of the temperature-dependent S~Doradus instability strip \citep{W89} and the constant temperature strip of 
 LBVs in outburst, as in \citet{S04}. The new {\it Gaia} DR2 distance brings the LBV W243 to the lower edge of the S~Doradus instability strip. YHGs and RSGs \citep{C05,M07} have a wide range of zero-age-main-sequence masses which could be due to errors in bolometric correction \citep{D18}, variations in reddening, or could be due to binaries. The single-star model tracks (blue) are from \citet{Bd11}. The evolutionary tracks do not reproduce the WR phases.  The LBV W243 has an inferred mass of 23.9$^{+2.8}_{-3.2}\rm{M_{\odot}}$, an age of 7.6$^{+1.0}_{-0.9}$ Myr.  The brightest YHG has an inferred initial mass of around 28.6$^{+4.1}_{-4.8}\rm{M_\odot}$, an age of 5.5$^{+1.6}_{-0.5}$ Myr.}
\label{fig:hrd}
\end{figure*}

If we assume that LBV W243 is a representative of the cluster, then the age of the cluster is 7.6$^{+1.0}_{-0.9}$ Myr, the turn-off mass is 22.3$^{+2.2}_{-2.4}\rm{M_\odot}$ (down from 40M$_\odot$), and the mass of the most evolved stars is 28.6$^{+4.1}_{-4.8}\rm{M_\odot}$. Fig.~\ref{fig:hrd} shows that the new inferred luminosity brings the LBV W243 to the lower edge of the S~Doradus instability strip  \citep{S04}. However, Fig.~\ref{fig:hrd} also clearly shows that there are RSGs in Wd1 with implied initial masses below 20 M$_{\odot}$, well below the presumed turn-off mass even at the nearer distance, and implied ages of around 10 Myr.  This may suggest either uncertain bolometric corrections, a range of ages in Wd1, or may point to the influence of binary evolution on the evolved star population (see e.g. \citealt{beasor19}).

Most of the prior distances for Wd1 relied on measuring an apparent magnitude, assigning an absolute magnitude based upon the stellar type, and calculating the distance modulus.  However, Wd1 suffers from high extinction, with an inferred $A_V$ of about 11 mag \citep{C05,D16}.  The uncertainty in the reddening translates to highly uncertain true apparent magnitude estimates, and hence, highly uncertain luminosity-based distance estimates. With an independent geometric {\it Gaia} DR2 distance, the reddening and bolometric luminosities of cluster stars can be re-evaluated, although this is beyond the scope of our paper.

One final but important point concerns the total stellar mass of Wd1. This cluster has been discussed as potentially one of the most massive young star clusters in the Galaxy \citep{C05}.  However, in addition to lowering the luminosities of the evolved stars, lowering the cluster turn-off mass, and raising the cluster's age, the smaller distance from DR2 also lowers the total mass of the cluster.  The inferred very high total stellar mass of the cluster of $\sim$10$^5$ M$_{\odot}$ was derived by integrating down a Kroupa IMF from the turn-off mass of around 40 M$_{\odot}$, and by scaling relative to the number of observed evolved supergiant stars of initially 30-40 M$_{\odot}$.  If the revised DR2 distance lowers all luminosities by -0.58 dex, and hence the turn-off mass from 40 to 22 M$_{\odot}$ as noted above, then the expected relative number of evolved stars increases by a factor of $\sim$4 at the turn-off mass.  Normalizing to the observed number of evolved stars therefore lowers the extrapolated total stellar mass of the cluster by roughly the same amount, which would make Wd1's initial mass comparable to or less than the mass of the Arches cluster in the Galactic centre, although Wd1 would be significantly older \citep{k00,S02,H10}.

\section{Conclusion}\label{sec:consclusion}
We use {\it Gaia} DR2 parallax measurements and Bayesian inference to estimate the distance to the Westerlund 1 (Wd1) massive star cluster, as well as the distribution of field stars along the line of sight. We model both cluster stars and Galactic field stars, and we find that the cluster parallax is $0.35^{+0.07}_{-0.06}$ mas, which corresponds to a distance of $2.6^{+0.6}_{-0.4}$ kpc. 
The new distance represents the highest precision, 18\%, to Wd1 to date. Much of this precision is limited by the systematics such as parallax zero-point, which is included in the Bayesian model. However, the models are rough and conservative, and require improvement in the future. For example, rather than using one zero-point value, we consider a distribution of zero-points due to the observed variation of the parallax zero-point. we also consider different offsets for the field and cluster stars. This model is a rough estimate, and to further improve the parallax precision will either require better models for the systematics, or better calibration in subsequent data releases.

Wd1 has been discussed as potentially one of the most massive young star clusters in the Galaxy, but revising the distance to this one cluster reduces its total mass and increases its age, and may have profound consequences for stellar evolution theory. An improved distance can significantly narrow the precision on luminosity, mass, and age of the cluster, which provides constraints on the post-main-sequence evolution of cluster members. Based on the new {\it Gaia} distance, we infer turn-off mass of around 22 M$_{\odot}$, which implies that the progenitor mass of the magnetar CXO J164710.2–455216, and LBV W243 is a little bit above 22 M$_{\odot}$.

\section*{Acknowledgements}
\scriptsize
This work has made use of data from the European Space Agency (ESA) mission {\it Gaia} (\url{https://www.cosmos.esa.int/gaia}), processed by the Gaia Data Processing and Analysis Consortium (DPAC, \url{https://www.cosmos.esa.int/web/gaia/dpac/consortium}). Funding for the DPAC has been provided by national institutions, in particular the institutions participating in the Gaia Multilateral Agreement.
This project was developed in part at the 2018 Gaia Sprint, hosted by the eScience and DIRAC Institutes at the University of Washington, Seattle.
Support for MA and JWM was provided by the National Science Foundation under Grant No. 1313036.
Support for NS was provided by NSF awards AST-1312221 and AST-1515559, and by the National Aeronautics and Space Administration (NASA) through HST grant AR-14316 from the Space Telescope Science Institute, which is operated by AURA, Inc., under NASA contract NAS5-26555.

\bibliographystyle{mnras}
\bibliography{MA}
\label{lastpage}
\end{document}